# Design for tunable optofluidic optical coupler with large dynamic range

Xionggui Tang*, Fang Meng

Key Lab of Low Dimension Quantum Structures and Quantum Control of Ministry of Education, College of Physics and Information Science, Hunan Normal University, Changsha 410081, P. R. China

* Email: tangxg@hunnu.edu.cn

**Abstract**: A novel scheme for tunable optofluidic optical coupler is proposed, by using directional coupling waveguide structure and microfluidic channel with two tapers at end points. The normalized optical power at two output ports can be dynamically manipulated by controlling the refractive index of liquid mixture in microfluidic channel. The optical performance of the designed device is numerically investigated by employing the beam propagation method (BPM). The simulated results demonstrate that large dynamic range and low optical loss for both TE and TM mode can be easily achieved, and furthermore the dependence of polarization states and operation wavelength is very low in our designed device. In addition, the tunable optofluidic coupler has advantages including simple structure and large fabrication tolerance. Accordingly, our proposed device offers a new approach for manipulation of optical power output, which has wide potential application in optofluidic systems.

## 1. Introduction

Optofluidics, called as the marriage of photonics and microfluidics, is a rapidly developing research field, and has attracted increasing attention in recent years [1-2]. It provides a new opportunity for achieving novel function devices, which have wide potential applications in many areas such as chemistry analysis, biosensor, drug discovery and environmental monitoring. The optofluidic devices have unique advantages including high tunability, rapid prototyping and ease of reconfiguration [3]. The approaches for achieving tunability of optofluidic devices are generally realized by refractive index modulation, resulting in change of the phase, the amplitude, or the polarization of lightwave. In past several years, the different optofluidic devices have been proposed, which includes optofluidic laser, optical limiter, modulator, optical attenuator and optofluidic sensor, etc. [4-11], based on the interaction between mcirofluidic matter and light field at the microsacle. Currently, optofluidics is an emerging area, where many researchers are actively developing various optofluidic devices with novel functions and exploring their potential for applications in the optofluidic systems.

The optical coupler is an important optical component, which is widely used in photonic integrated systems as optical function devices, including optical switch, optical modulator, optical router. For these reported devices, the desired optical functions have been achieved, by generally using electro-optical effect, or thermo-optical effect etc. However, these optical couplers are inappropriate for application in optofluidic systems. Therefore, it is very essential to exploit the tunable optofluidic coupler used in optofluidic systems. In the past several years, some approaches for optofluidic couplers have been demonstrated. In 2012, Marius Vieweg et al. presented a tunable optofluidic optical coupler based on Kerr nonlinearity effect and thermo-optics effect by using photonic crystal fiber filled with liquid CCl4, which has advantages such as rapid tunability speed and large dynamic range [12]. In the following, Heming Wei et al. proposed an optofluidic optical coupler based on evanescent field coupling between photonic crystal fiber and single mode fiber, used as a refractive index sensor [13]. However, these schemes mentioned above are based on fiber structures, so they can't be utilized in optofluidic chip-on-lab platforms based on waveguide structures. In addition, Mehdi Hosseinpour designed an optofluidic optical coupler by employing the photonic crystal structure filled with different liquids [14]. However, its optical performance is generally sensitive to the structure parameters, operation wavelength and polarization, which limits its potential applications in optofluidic systems. Later, Guomin Jiang provided an approach for realizing the optofluidic optical coupler, which consists of waveguide with 45o mirrors and the perpendicular microfluidic channel, and it is used as a switching coupler in card-to-backplane optical interconnection [15]. In this case, however, the proposed device has complex structure, which is relatively difficult to be precisely fabricated and be accurately controlled, resulting in high cost.

In this work, we propose a novel scheme for tunable optofluidic optical coupler, in which the directional coupling waveguide structures are employed, and microfluidic channel with two tapers at end points is introduced. The liquid mixture with controllable refractive index is injected into the microfluidic channel, used as a part of the upper cladding of waveguide at modulation region. The optical power output can be dynamically changed by controlling the refractive index of the liquid mixture. Through numerical simulation, the proposed device exhibits excellent optical performance, which includes large dynamic range, low optical loss, low dependence of operation wavelength and polarization state, so it is strongly preferred in optofluidic systems. Consequently, we believe that the optofluidic optical coupler has wide potential application in optofluidic systems. For instance, it could be used as refractive index sensor, optical attenuator, optical power splitter in optofluidic platform.

## 2. Structure design

The schematic structure of tunable optofluidic optical coupler is depicted in Fig. 1, in which Fig. 1(a) is its top view, and Fig. 1(b) and (c) denote the cross-section view at the position $S_1S_2$ and $P_1P_2$, respectively. The designed configuration includes the waveguide structure and microfluidic channel structure. For the former, it consists of one input port and two output ports, in which the straight waveguide and bending waveguide are together employed in this configuration. In this structure, the waveguide structures are composed of top cladding, the core and the bottom cladding, whose refractive indices are $n_1$, $n_2$ and $n_3$, respectively. Correspondingly, the slab height, rib height and the core width are denoted by $d$, $h$ and $w$, respectively. At the coupling region, the two waveguides are parallel, and their gap width is equal to $g$. For the proposed device, the microfluidic channel with taper shape at two end points are introduced to be filled with liquid mixture as the top cladding, whose refractive index can be adjusted by manipulating the concentration of liquid mixture.

The operation mechanism of the proposed device is given in the following. The liquid mixture with controllable refractive index is injected at input port $Q_1$ by using syring actuated by pump, and then it flows across the microfluidic channel. Afterwards, the liquid flows out at the output $Q_2$. Seen from Fig. (1), the liquid mixture at coupling region is employed as its upper cladding. The optical signal is coupled into waveguide input by using tapered lens fiber, and it propagates forward along waveguide. At the coupling region, the coupling mechanism is based on evanescent field coupling in our designed configuration, and its coupling efficiency is commonly determined by structure parameters and the refractive index profile. As a result, the optical power at two output ports A and B can be dynamically manipulated by changing the refractive index of the liquid mixture. As is well known, the liquid generally has high tunability of refractive index by controlling the concentration or temperature of liquid mixture. In our layout, the proposed optofluidic coupler has potential for achieving the large dynamic range of optical splitting ratio.

As mentioned above, the dynamic range of optical splitting ratio is a crucial parameter used for evaluating the optical performance of the optofluidic optical coupler. For TE and TM mode input, the dynamic ranges are defined below, respectively,

$$\text{DR}^{TE} = Max[10\log(R^{TE})] - Min[10\log(R^{TE}), \tag{1}$$

$$\text{DR}^{TM} = Max[10\log(R^{TM})] - Min[10\log(R^{TM}), \tag{2}$$

where *Max* and *Min* stand for the maximal and minimum value in the bracket, respectively. Here, $R^{TE}$ and $R^{TM}$ denote optical splitting ratio for TE and TM mode input, respectively, which are given as follows,

$$R^{TE} = P_A^{TE}/P_B^{TE}, \tag{3}$$

$$R^{TM} = P_A^{TM}/P_B^{TM}, \tag{4}$$

where $P_A^{TE}$ and $P_B^{TE}$ are the optical power of TE mode at output ports A and B, respectively, and $P_A^{TM}$ and $P_B^{TM}$ represent its optical power of TM mode at output ports A and B, respectively. In addition, its optical excess loss for TE and TM modes is written as,

$$L^{TE} = -10 \log(P_A^{TE} + P_B^{TE})/P_{in}^{TE}], \tag{5}$$

$$L^{TE} = -10 \log(P_A^{TE} + P_B^{TE})/P_{in}^{TE}], \tag{6}$$

where $P_{in}^{TE}$ and $P_{in}^{TM}$ indicate the optical power of TE and TM mode input, respectively. Seen from Fig. 1, the tunable optofluidic optical coupler is a 3-D waveguide structure. In this work, the effective index method is used to change the 3-D waveguide into the 2-D one, in order to effectively reduce computation cost. Then, the numerical simulation is performed to investigate the optical performance of the optofluidic optical coupler, by using Beam Propagation Method (BPM) [16].

3. Numerical Simulation and Analysis

In numerical simulation, the operation wavelength is assumed to be 1550nm. The polymer materials UV-15, SU-8 and PMMA are chosen as the bottom cladding, the core and the top cladding of waveguide structure, respectively, whose refractive indices are 1.500, 1.575 and 1.492, respectively. For simplicity, the refractive index of the liquid varies from 1.490 to 1.535, by controlling the concentration of liquid mixture. The related other parameters are given as follows. The waveguide width, slab thickness and rib height are 5.0 μm, 0.9 μm and 0.7 μm, respectively. At the coupling region, the gap width is 3.0 μm, and coupling length is 640μm. For TE mode input, the effective refractive indices in the gray and black region are 1.5438 and 1.5256, respectively. For TM mode input, however, the effective refractive indices in the gray and black region are 1.5423 and 1.5234, respectively.

The liquid mixture, which is injected into the microfluidic channel, acts as the upper cladding of waveguide structure in the coupling region, and its refractive index can be controlled by manipulating its concentration of liquid mixture. Fig. 2 (a) and (b) present the normalized optical power at output ports A and B, when the TE and TM mode are coupled into the input port, respectively. The numerical results demonstrate that the normalized optical power at output port A declines, but the normalized optical power at output port B rises, while the refractive index of liquid mixture varies from 1.49 to 1.535. More importantly, the optical coupling properties for TE mode input are approximately same with those for TM mode input. It exhibits that the proposed optofluidic coupling has very low dependence on the polarization states, which is highly desired in potential application for integrated optofluidic systems. In addition, the corresponding dynamic range for both TE and TM mode are above 45 dB, according to the numerical results in Fig. 2. Particularly, their optical excess losses are smaller than 0.06 dB. It exhibits good optical performance, including large dynamic range and low optical excess loss. Further, the optical field distribution for TE mode input in the optical coupler is given in Fig. 3 (a) and (b), while the refractive indices of liquid mixture are equal to 1.490 and 1.530, respectively. The simulated results show that the magnitude of the coupled optical power can be effectively controlled by changing refractive index of liquid, resulting from the change of coupling coefficient.

Broadly speaking, spectrum width of laser source is about several to several decade nanometers. In the following, the optical performance dependence on operation wavelength is numerically investigated. In this case, the optical source is supposed to have spectrum width from 1500 nm to 1600 nm. It is worthy to be mentioned that the dispersion of refractive index of polymer materials and the liquids are generally weak in near infrared region. For example, the refractive index of PMMA changes from 1.49256 to 1.49237 in the wavelength scope from 1500 nm to 1600 nm. While the optical wavelength varies from 1500 nm to 1600 nm, normalized optical power at output port A and B, for the TE and TM mode, are given in Fig. 4 (a) and (b), respectively, which are depicted by solid and dashed lines, respectively. Here, the refractive index of liquid mixture is assumed to be 1.510, and the other parameters are same with those in Fig. 2. As seen from Fig. 4, the variation of optical power in the wavelength range of 100 nm is very small, for both TE and TM mode. It shows that the wavelength dependence of the proposed device is very weak, which is highly preferred in potential application in the integrated photonic systems.

In the fabrication process, there always exists fabrication error, which always has a negative effect on the optical performance of the proposed device. Here, the optical performance affected by fabrication error of waveguide width, gap distance, rib and slab thickness are numerically investigated in the following. In this case, the optical signal is assumed to be TE mode, since the polarization dependence is very weak. The refractive index of liquid mixture is supposed to be 1.510, and the other parameters are same to those in the Fig. 3 (a). The simulated results are given in Fig. 5 (a), (b), (c) and (d), respectively. In Fig. 5 (a), the normalized optical power at output port A and B changes slightly while the deviation of waveguide width varies from -0.5μm to +0.5μm. Seen from Fig. 5 (b), the normalized optical power at output port A increases slowly, but the normalized optical power at output port B gradually decreases while the deviation of gap width changes from -0.5μm to +0.5μm, respectively. In Fig. 5 (c) and (d), the optical power at output port A and B are almost unchanged while the deviation of rib height and slab thickness change from -0.1μm to +0.1μm. Consequently, the fabrication error effect on the optical performance is small, which can be ignored, due to fact that the optical power output can be dynamically manipulated by controlling the refractive index of liquid mixture. According to our numerical simulation, it suggests that our proposed device has relatively large fabrication tolerance, which is useful for fabrication of device.

## 4. Conclusions

In this work, a novel tunable optofluidic optical coupler has been proposed, by utilizing directional coupling waveguide structure, and microfluidic channel with two tapers at end point. The refractive index of liquid mixture in microfluidic channel can be controlled to achieve the tunability of optical power output. The simulated results indicate that our designed optofluidic coupler has excellent optical performance, which includes large dynamic range, low optical loss, and weak dependence of wavelength and polarization states. Additionally, the fabrication error effect on the optical performance has also been briefly analyzed, and the simulated results reveal that the device has a relatively large fabrication tolerance. Accordingly, our proposed device has potential for application in the integrated optofluidic system, which can be used as optofluidic functional device such as refractive index sensor or biophotonic sensor in optofluidic platform.

**Acknowledgments**

This work is supported in part by the Natural Science Foundation of Hunan Province in China under grant 2016JJ2087. The authors also acknowledge support by the Youth Science and Technology Foundation of Hunan Normal University and Foundation of China Scholarship Council.

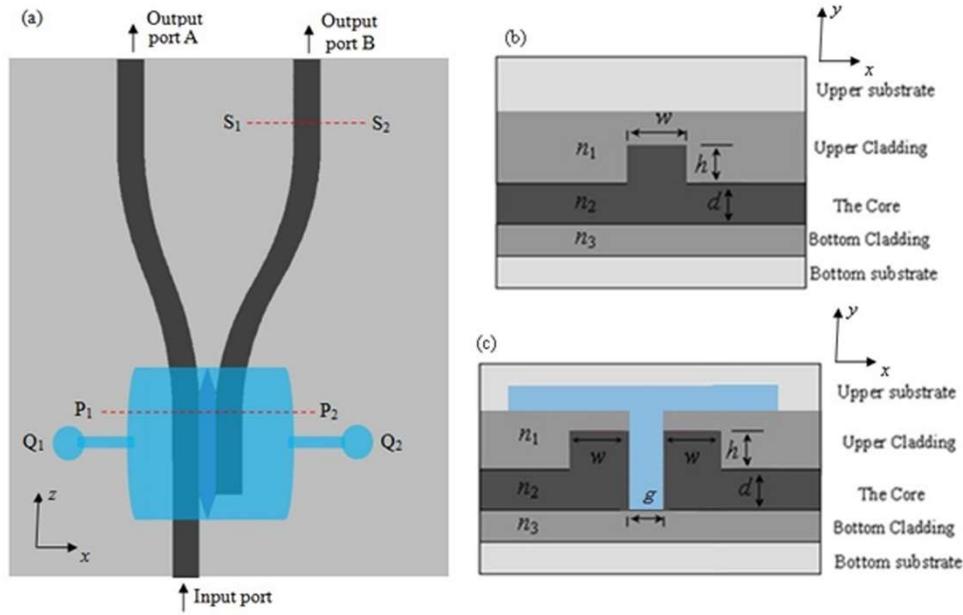

**Fig. 1.** Schematic diagram of tunable optofluidic coupler, (a) its top view, (b) and (c) its cross-section views of waveguide at the position $S_1S_2$ and $P_1P_2$, respectively.

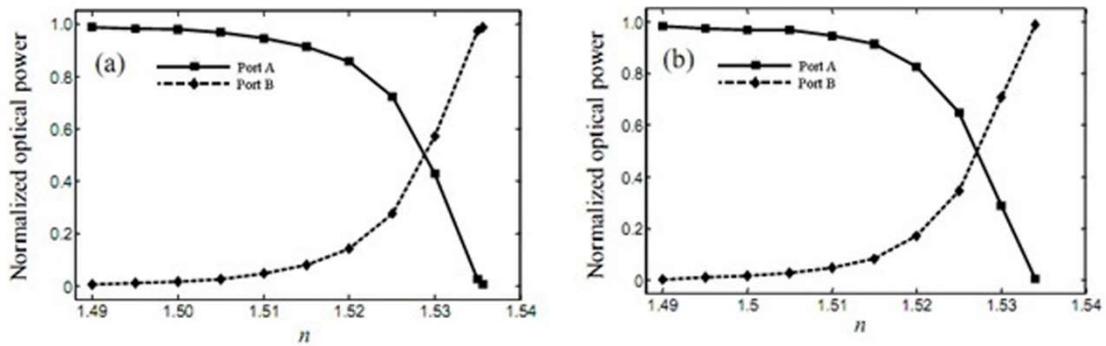

**Fig. 2.** The normalized optical power at output ports which varies with the refractive index of liquid mixture: (a) TE mode and (b) TM mode.

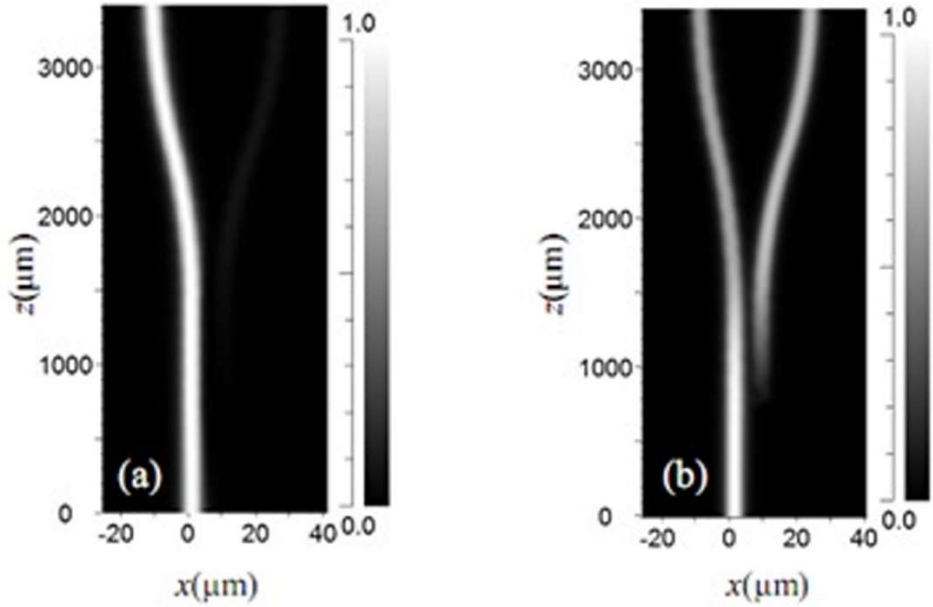

**Fig. 3.** The optical field distribution of TE mode when the refractive index of liquid mixture is (a) 1.490 and (b) 1.530, respectively.

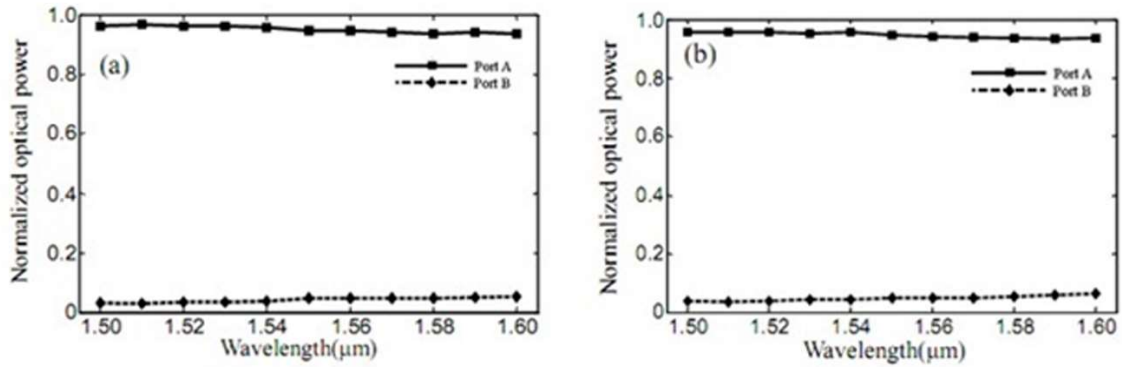

**Fig. 4.** The normalized optical power at output port A and B as a function of operation wavelength: (a) TE mode and (b) TM mode.

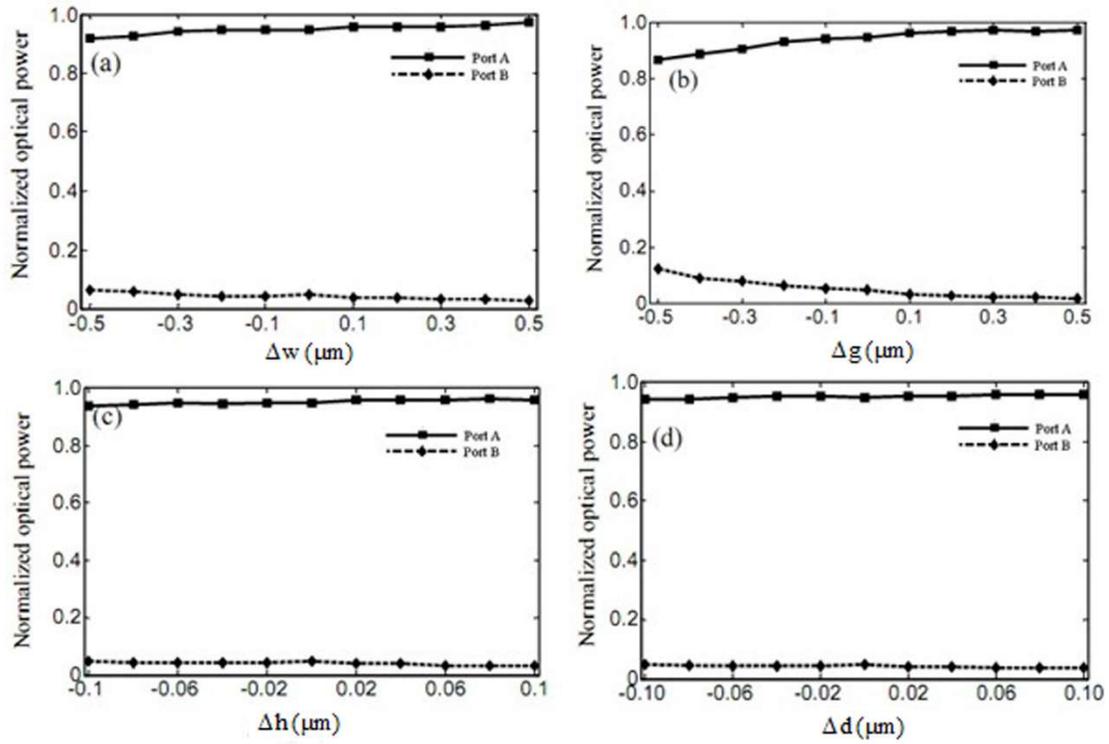

**Fig. 5.** The normalized optical power output varying with different fabrication deviation of (a) waveguide width, (b) gap width, (c) rib thickness and (d) slab thickness.